\documentclass[preprint,12pt,authoryear]{elsarticle}

\usepackage{amssymb}
\usepackage{amsmath}
\usepackage{multicol}
\usepackage{multirow}

\begin{document}

\begin{frontmatter}

\title{Citizens' Contentment with e-Government Solutions and Services in Saudi Arabia}

\author{Mohammed O. Alannsary}
\affiliation{organization={Department of Digital Transformation Programs, The Institute of Public Administration},
            city={Jeddah},
            country={Saudi Arabia}}
\ead{alannsary@hotmail.com}

\begin{abstract}
Governments around the world have worked tirelessly to develop technological solutions, on the one hand to better serve their citizens and, on the other hand, to advance in the United Nations Electronic Government Development Index (EGDI). Thus, it is crucial to assess e-government solutions and services from different aspects. This study evaluates e-government solutions and services based on user expectations in four aspects: general satisfaction, satisfaction with features, trust, and pleasure. In this study, a questionnaire was developed to allow the evaluation of e-government solutions and services in Saudi Arabia, and could also be utilized to evaluate e-government in any other nation. The study included 276 valid participants, while the required sample size was calculated using a standard sample size estimation formula for large populations (95\% confidence level, 5\% margin of error). In addition, descriptive analysis was used to analyze participant responses. The results showed that e-government services in Saudi Arabia achieved a level of citizen contentment that is consistent with its score in the EGDI published in the year 2024.
\end{abstract}


\end{frontmatter}

\section{Introduction}
\label{Intro}

The United Nations Department of Economic and Social Affairs has been publishing the UN e-government survey since 2001 \cite{united_nations_department_of_economic_and_social_affairs_united_2024}. The purpose of the survey was to evaluate the state of e-government initiatives across all United Nations (UN) members. The results presented serve as a benchmark by which UN members can measure their progress in e-government related aspects, and work on improving it.

The primary function of the Digital Government Authority (DGA) \cite{government_of_saudi_arabia_digital_2023}, the responsible government body in charge of the electronic government program in Saudi Arabia, is to clear the path for government agencies to efficiently and effectively deliver their services digitally. DGA is also in charge of creating laws, rules, and regulations that are necessary to enhance Saudi Arabia's digital transformation. Additionally, DGA monitors, assesses, and measures the digital capabilities and performance of Saudi government entities.

Government agencies have been urged to get their services ready for electronic access since the beginning by DGA and its predecessor, the Saudi Arabian e-government program "Yesser". These organizations did this by requiring the public to use the agency's portal and mobile application instead of going to offices in person to receive services. High-quality software solutions must be created for the portals and mobile applications in order to properly comply with such standards and make proper use of the services.

Information systems facilitate the sharing of data and information across linked parties in e-government \cite{hofmann_identifying_2012}. As a result, software is used by most e-government initiatives globally to deliver their solutions and services.

There is a significant relationship between quality and customer satisfaction or contentment, Juran \cite{juran1999think} explained that there are multiple meanings of "Quality", one of which refers to the aspects of any product that suits customers’ needs and eventually gain their satisfaction. Thus, e-government solutions and services that are of high quality need to suit customers' needs and eventually lead to their satisfaction.

In Saudi Arabia and globally, software has become a necessary tool. For tasks like scheduling appointments, completing government forms, and purchasing products, humans typically rely on software. Because of this dependence, it is necessary that these software solutions be of a high quality, which calls for analyzing and judging the program from a quality standpoint to ensure that it will function as intended when required. As a result, the International Organization for Standardization created a number of software engineering-related standards, including ISO/IEC 25010. Systems and software engineering, systems and software quality requirements and evaluation (SQuaRE), and system and software quality models are all covered by ISO/IEC 25010 \cite{ISO_systems_2011}.

When put into practice, the ISO/IEC 25010 standard is crucial for ensuring software quality \cite{ISO_systems_2011}. Furthermore, the ISO/IEC 25010 specifies two models: the quality in use model (QinU) and the product quality model. The product quality model comprises eight characteristics that are further divided into sub-characteristics that describe the system's dynamic qualities as well as the software's static features. The five characteristics of the QinU model, some of which are further subdivided into sub-characteristics, are focused on how users interact with the software in a particular situation. In order to accomplish the goal of this study, the Satisfaction characteristic is used partially.

A component of the SQuaRE series of international standards, ISO/IEC 25020: Quality Measurement Division \cite{ISO_quality_2011-1} focuses on the measurement reference model and guide, offering a road map for creating and defining quality measures. One of its elements, ISO/IEC 25022: Measurement of quality in use publication \cite{ISO_measurement_2011}, is concerned with QinU measures, which explain the measure for each characteristic and sub-characteristic. Each measure is specified using a measure ID, measure name, measure description, measurement function, and data collection method.

Because e-government solutions and services depend on software, this study develops a questionnaire using the ISO/IEC 25010 paradigm. In addition, it helps close the gap mentioned in \cite{santa_role_2019}, which is the lack of research on the impact of trust on user satisfaction in Saudi Arabia. The questionnaire assesses citizens' contentment with e-government solutions and services in Saudi Arabia. The satisfaction characteristic of the ISO/IEC 25010 QinU model is linked to a subgroup of sub-characteristics that are used to create the questionnaire's questions. 

The Questionnaire used in the study makes it different from others for several reasons, its questions are based on the characteristics of the ISO/IEC 25010 QinU model to guarantee that the right software quality aspects are studied, it enables evaluating e-government solutions and services from the viewpoint of users, and it functions as an independent instrument for measuring, assessing, and evaluating e-government solutions and services. In addition, the questionnaire is constructed to address four questions:
\begin{enumerate}
    \item When utilized in the designated context, do e-government services typically meet the needs of beneficiaries?
    \item Are users satisfied with the features of e-government services?
    \item Are e-government services trusted by users?
    \item Do consumers find e-government services enjoyable to use?
\end{enumerate}

The aforementioned questions contribute to the main objective of the study, which is answering the question: \textbf{How satisfied are Saudi Arabian citizens with e-government solutions and services?}

The remaining sections of this study are organized as follows: Background information and related work are presented in the next section. The questionnaire is described in Section 3. The results analysis is covered in Section 4. The conclusion and prospective are presented in Section 5. Lastly, a sample of the questionnaire is displayed in Appendix A.


\section{Background and Related Work}
\label{BG}
e-government can be defined in a number of ways, but the most widely accepted definition is based on e-government standards, which suggest that information systems are used to allow related parties to share information and data \cite{hofmann_identifying_2012}. "The use of IT in government operations, including its effects on public service delivery, citizens' satisfaction, and democratic standards" \cite{arias_digital_2018} is an alternate definition. Since 1999, several EU nations have incorporated the deployment of e-government services into their strategic plans \cite{ziemba_assessing_2014}. Such initiatives are noteworthy.

As was already noted, DGA is in charge of Saudi Arabia's e-government program. Previously, the program was called Yesser, and it was created in 2005 as a collaborative effort between the Ministry of Finance (MoF) \cite{government_of_saudi_arabia_mof}, the Ministry of Communications and Information Technology (MCIT) \cite{government_of_saudi_arabia_mcit}, and the Communications, Space \& Technology Commission (CSTC) \cite{government_of_saudi_arabia_cstc}, previously known as Communication and Information Technology Commission (CITC). All Yesser's projects and efforts were transferred to the DGA after its creation by a Saudi Arabian royal decree in March 2021. Currently DGA is overseen by MCIT \cite{alannsary_adopting_2019}. 

In the most recent survey of the E-Government Development Index (EGDI) released in 2024 \cite{united_nations_department_of_economic_and_social_affairs_united_2024}, Saudi Arabia was rated sixth out of all UN countries with an EGDI of 0.9602, and in the “Very High” group. Saudi Arabia was in the "Very High" group with a rating class of V2 in 2022 and 2020, according to the surveys issued in those years. Its EGDI score was 0.8539 in 2022, and 0.7991 in 2020, placing Saudi Arabia in thirty-first in 2022 and forty-third in 2020. 

Quality definitions, according to Alannsary \cite{alannsary_quality_2016}, are intricate and sometimes perplexing. Such a definition involves a number of perspectives, including the perspectives of the users and the specifications. Juran \cite{juran_juran_1988} employs the widely accepted concept of quality, which is "fitness for intended use." As a result, stakeholders and people may hold divergent opinions regarding different facets and meanings of software quality \cite{tian_quality-evaluation_2004}.

Quality is both a significant factor and a crucial part of the whole picture when evaluating products and services \cite{traetteberg_quality_2020}. It is one of the most crucial indicators for determining whether a software product is successful or not, according to experts and practitioners \cite{blum_software_1992, ghezzi_fundamentals_2002, humphrey_managing_1989, tian_quality_2007, von_mayrhauser_software_1990}. Software quality, like other products, can be described in a variety of ways. This is because stakeholders have varying expectations \cite{tian_quality-evaluation_2004}.

The UTAUT model \cite{venkatesh2003user} offers a theoretical framework for understanding how QinU dimensions of the ISO/IEC 25010 model influence the adoption of e-government. In particular, performance expectancy (usefulness) and effort expectancy (ease-of-use) serve as mediators in the relationship between system quality and user satisfaction.

Ramadhan and Pribadi \cite{ramadhan2024building} analyzed variables such as digital literacy, trust in government, its perceived usefulness, and the quality of electronic services using a quantitative research design and Structural Equation Modeling (SEM). The authors found that citizen satisfaction is affected by digital literacy and the electronic service quality.

Muara et al. \cite{muaramobile} suggested a model to assess the effectiveness of mobile government (m-government) projects from the viewpoint of the public, taking into account how they view the value of government services. using a survey with 233 participants (106 Brazilians and 127 Mozambicans) and using Structural Equation Modeling (SEM) to analyze the results. The study looks at how citizens' satisfaction with m-government services is affected by perceived value. The results show that citizens' satisfaction with m-government systems is significantly influenced by perceived value, and that this relationship is moderated by facilitating factors.

Idriss and Khalifa \cite{khalifa2023intermediary} suggested a model to gauge satisfaction with the Directorate General of Taxes' online tax services (SIMPL). The model is based on the DeLone and McLean models, the TAM model, and Mayer et al. \cite{mayer1995integrative}. Motivated by the TAM model, the model considers perceived utility. A sample of 168 SIMPL taxpayers from all over Morocco was used to test the model. The hypotheses were tested using the Structural Equation Modeling (SEM) approach. In their work, they found that perceived usefulness and trust both impact user contentment.

Pramuditha et al. \cite{pramudithaexploring} investigated the impact of e-government service quality on citizen satisfaction and trust. 379 users of an online platform participated in a survey as part of a quantitative research strategy. The results show that citizen contentment is significantly positively impacted by the quality of e-government services, with information quality appearing as the most important metric. In addition, they found that citizen contentment has a favorable impact on citizen trust.

Taufiqurokhman et al. \cite{taufiqurokhman2024impact} investigated the degree to which public satisfaction can be impacted by the quality of e-services. Using primary data sources and a quantitative approach, the research method used random sampling as its sampling strategy. The study participants were residents who used government digital public service platforms. In their work, the sample size included 262 participants. Public satisfaction, public trust, and e-service quality are the variables examined in the study. The research findings demonstrated that both public trust and public satisfaction are significantly impacted by the quality of e-services. In addition, the investigation refutes the notion that the relationship between e-service quality and public satisfaction is mediated by public trust.

Ziemba et al. \cite{ziemba_assessing_2014} claimed that e-government websites must be of a high quality in order to be successfully accepted. Consequently, they put forward a framework and employed it to assess the quality of three Polish e-government portals. The framework is based on the ISO/IEC 25010 standard, the product quality model, in particular. Although data analysis was unrestricted, the authors indicate that the limited sample size was seen as a restriction of their investigation. Furthermore, the framework was only examined and validated by e-government portal staff.

Alghamdi et al. \cite{alghamdi_organizational_2014} examined Saudi Arabia's organizational preparedness for e-government. To examine the attitudes, strategies, successes, and challenges related to e-government, top e-government leaders were interviewed. In addition, to evaluate the organizational readiness in e-government, a novel model that concentrated on public sector organizations was presented in the study, in contrast to previous models. Additional elements like process and strategy were also covered by the model.

Makki and Alqahtani \cite{makki2022modeling} investigated the obstacles of adopting e-government in Saudi Arabia and created a model to classify the thirteen obstacles found according to their reliability and strength of motivation, exposing interrelationships between them on multiple levels through the use of Interpretive Structural Modeling (ISM). A deeper comprehension of the contextual relationships between the barriers is one of the model's consequences, which will help sustain current implementation successes and create the possibility of new opportunities in the future.

Saudi Arabia's e-government has been examined from different aspects, Al-Sakran and Alsudairi \cite{al-sakran_usability_2021} examined the usability and accessibility of mobile websites across a number of public sectors. Hussain \cite{hussain_factors_2020} looked at the elements that affected Saudi Arabian citizens adoption of e-government. Aloboud et al. \cite{aloboud_evaluating_2020} used the Web Content Accessibility Guidelines and Nilsson's 10 Heuristics to examine the usability and accessibility of e-government websites. A number of suggestions were made to address issues with usability and accessibility. Alharbi et al. \cite{alharbi_overview_2020} concentrated on the elements that encourage the use of mobile government solutions by Saudi Arabian citizens, particularly its success. Almukhlifi et al. \cite{almukhlifi_e-government_2018} examined, from the point of view of the population, the moderating impact of Wastta, a Saudi cultural practice, on the adoption of e-government. Santa et al. \cite{santa_role_2019} investigated how user satisfaction, the efficiency of e-government systems, and operational effectiveness are impacted both directly and indirectly by trust in online services.

Analysis of the literature indicates that there has been no investigation or assessment of Saudi Arabian citizens' contentment with e-government solutions and services.


\section{Questionnaire}
\label{Quest}

The QinU model's characteristics and sub-characteristics that use questionnaires as a data collection tool form the foundation of the questionnaire. The \textbf{Satisfactory} characteristic has sub-characteristics that employ a questionnaire as a means of gathering data, in accordance with ISO/IEC 25022: measurement of quality in use \cite{ISO_measurement_2011}. Furthermore, because the \textbf{comfort} sub-characteristic focuses on bodily comfort, which is outside the scope of this study, it is not included in the questionnaire. The \textbf{Satisfaction} characteristics of the QinU model is used in the questionnaire including the \textbf{Usefulness, Trust, and Pleasure} sub-characteristics. Additionally, a subset of the ISO/IEC 25022 Satisfaction characteristic measures that were used in the current study are named and described next: 

\begin{description}
    \item[General] Overall satisfaction: The overall satisfaction of the user.
    \item[Usefulness] Satisfaction with features: The satisfaction of the user with specific system features.
    \item[Trust] User trust: The extent to which the user trusts the system.
    \item[Pleasure] User pleasure: The extent to which the user obtains pleasure compared to the average for this type of system.
\end{description}
    
In this research, the notion of "satisfaction" is regarded as a multifaceted construct, encompassing both objective and subjective components. It comprises task completion (i.e., outcome satisfaction), usability and clarity of service (i.e., process satisfaction), as well as user emotional responses like pride or frustration (i.e., emotional satisfaction). This comprehensive perspective is consistent with the QinU framework outlined in ISO/IEC 25022 and facilitates a more detailed assessment of user experiences with e-government services.

\begin{figure}[t]
\centering
\includegraphics{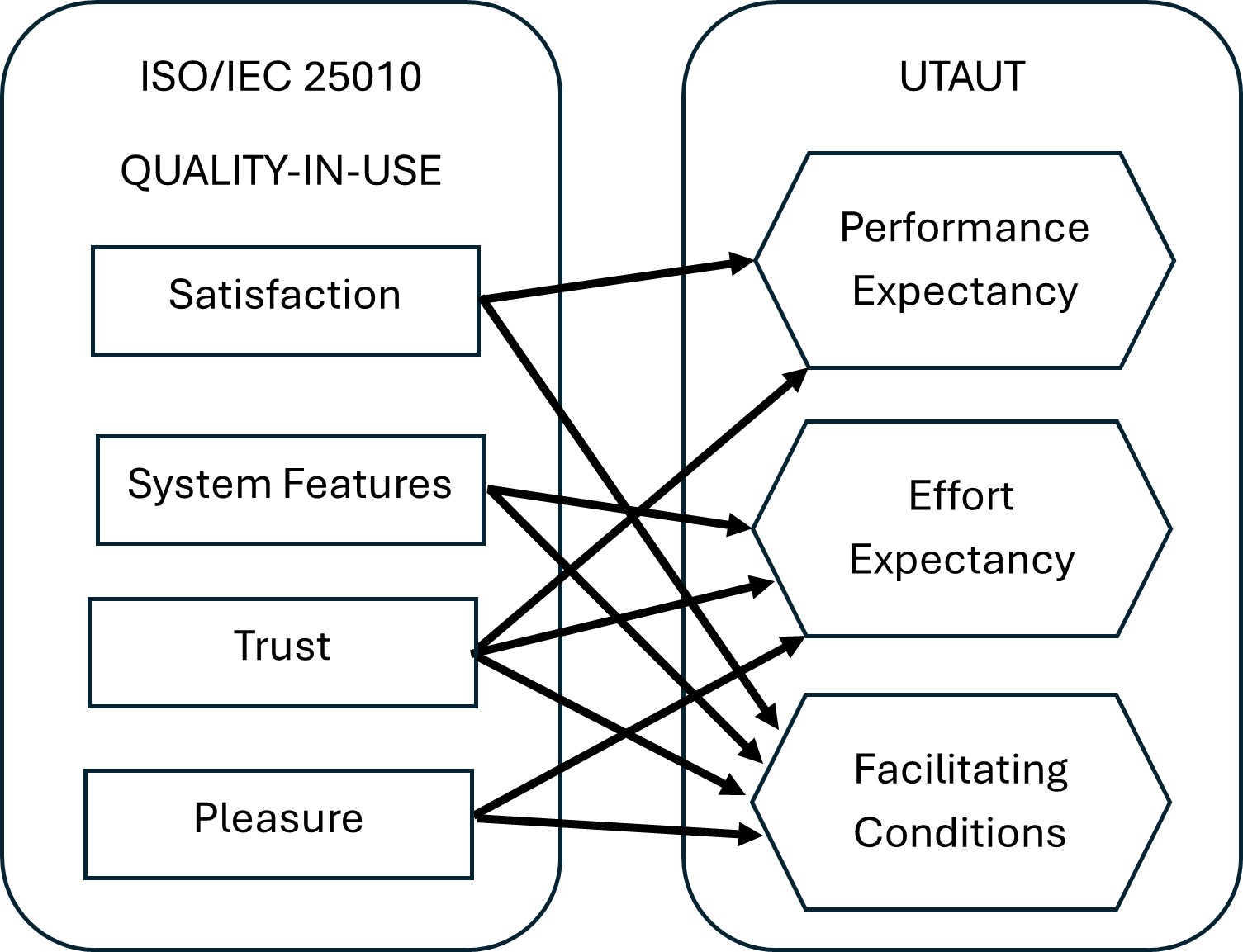}
\caption{Theoretical Framework Linking ISO/IEC 25010 QinU Dimensions(left) to UTAUT Constructs (right), illustrating how standardized quality metrics relate to behavioral drivers of technology adoption.}
\label{fig1}
\end{figure}

The identified patterns of user satisfaction correspond with the fundamental elements of the Unified Theory of Acceptance and Use of Technology (UTAUT) \cite{venkatesh2003user}. Elevated scores for usability and functional transparency indicate the impact of \textbf{effort expectancy}, whereas the perceived reliability and benefit of task completion reinforce \textbf{performance expectancy}. Furthermore, apprehensions regarding data privacy and fluctuating emotional involvement among specific demographics may be associated with \textbf{facilitating conditions} and \textbf{social influence}, especially within the context of government services. Although the research did not conduct statistical tests on UTAUT relationships, the model serves as a valuable framework for comprehending the variations in user trust and satisfaction. Figure \ref{fig1} demonstrates the manner in which the framework of the study connects ISO quality metrics (such as satisfaction) with the acceptance drivers of UTAUT (for instance, effort expectancy), thereby providing a theory-based perspective for the evaluation of e-government.

There are twenty-two closed-ended questions in the questionnaire. Basic information, e-government service and beneficiary needs, and satisfaction with features of e-government service are the three main areas into which these questions were divided. Additionally, a question was utilized to verify participant answers and exclude out those that were inadequate. The questionnaire contains a unique series of questions for each of the aforementioned factors: general satisfaction, satisfaction with features, trust, and pleasure as shown below. The questionnaire and all possible responses to its questions is provided in Appendix A.

\begin{description}
    \item [General satisfaction:] questions (6, 13, 15, 16, and 17).
    \item [Satisfaction with features:] questions (8, 9, 10, 11, 12, 18, 19, and 20).
    \item [Trust:] question (21).
    \item [Pleasure:] question (22).
\end{description}

It is important to note that the preliminary instrument employed in \cite{de_andrade_soares_evaluating_2019} served as the basis for the questions about general satisfaction and satisfaction with the features. Additionally, in accordance with the recommendations established in ISO/IEC 25022: measurement of quality in use, questions pertaining to the pleasure and trust aspects were based on those proposed and supplied in \cite{jian_foundations_2000, hassenzahl_inference_2010, watson_development_1988}. Furthermore, ten anticipated participants, five experts and five members of the public, were given the questionnaire to assess its validity, precision, and application. The questionnaire was modified according to their notes and remarks.

\subsection{Approach}

Since the majority of e-government solutions and services are offered to a large number of beneficiaries, a large population is the focus of any study that evaluates citizen satisfaction. All Saudi citizens and non-citizens who are in Saudi Arabia with a valid visa are included in the study's population. Because of the size of the population, it is difficult to get comments and input from the willing participants. The questionnaire was created and then disseminated online to gather answers in order to get past this obstacle. Therefore, it is recommended to compute the sample size using the basic random sampling technique with a 95\% confidence level and a 5\% margin of error.

\subsection{Population and Sample Size}

Saudi Arabia's General Authority for Statistics \cite{saudi_arabian_government_general_nodate} published its 2022 annual report: "Population Report" The report estimates that there are 32,176,224 individuals in Saudi Arabia overall, and that there are 22,925,763 people who are 18 years of age or older. Saudi Arabian laws and regulations grant everyone who is at least eighteen years old access to e-government services. Thus, the population of this study is 22,925,763. Due to the magnitude of the population, Simple random sampling was employed. The following formula is used to determine the sample size:

\begin{equation}n=(\frac{Z_{\alpha/2} ^2 * P(1-p)}{E^2})\label{eq1}\end{equation}

\begin{equation}n=(\frac{1.96_{\alpha/2} ^2 * 0.5(1-0.5)}{0.05^2}) \simeq 385\label{eq2} \end{equation}
where:
\begin{itemize}
    \item[] $n$ represents the simple random sample size,
    \item[] $Z_{\alpha/2} ^2$ represents the standard normal deviation (95\% confidence level), and
    \item[] $E$ represents the margin of error (5\%).
\end{itemize}

\subsection{Questionnaire Distribution}
The questionnaire was distributed to 500 people, yielding 402 responses (an 80.4\% response rate), which was deemed sufficient. Out of the responses that were received, only 276 were found valid. The remaining 126 responses were eliminated due to being insufficient, contradictory, or inconsistent. For example, if a participant responds by being "excited" and "jittery" at the same time when asked about their feeling when logging into an e-government service, the respondent's response will be disqualified.


\section{Analysis and Discussion of Results}
\label{analysis}

In this section, descriptive analysis is used to analyze questionnaire responses. Questions are linked to generate groups: the first group helps compile the participants' basic information, while the remaining four groups provide insight into the four aspects that are main pillars in this study and contribute to quality in general.

\subsection{Basic Information}

Questionnaire participants were of different age categories as shown in Table \ref{table5}: 34.4\% of participants were 30-39 years old, 28.6\% were 40-49 years old, and 16.7\% were 50-59 years old. Participants that were from these three categories witnessed the inception of the internet, and are thus more familiar with it. They are also responsible family members and need to use e-government services to fulfill their family obligations and needs. Table \ref{table6} display population age distribution as per The Saudi Arabian General Authority of Statistics. The 30 to 39 age group represents 32.09\%, the 40 to 49 age group represents 19.54\%, and the 50 to 59 age group represents 10.40\%. For the above mentioned three age groups, the percentage difference between the sample and the actual population is between 2.31\% and 8.86\%. This implies that the sample represents the population to an acceptable degree.

\begin{table} [h] 
\caption{Participant's age}
\centering
{\begin{tabular}{lcc}
        \hline
        Age & Count &	Percentage \\
        \hline
        Below 20 &	2&	0.7\% \\
        20 to 29&	18&	6.5\% \\
        30 to 39&	95&	34.4\% \\
        40 to 49&	79&	28.6\% \\
        50 to 59&	46&	16.7\% \\
        60 or above&	36&	13\% \\
        Total&	276&	100\% \\
        \hline
\end{tabular}}
\label{table5}
\end{table}

\begin{table} [h] 
\caption{Population age}
\centering
{\begin{tabular}{lcc}
        \hline
        Age & Count &	Percentage \\
        \hline
         18 to 19 &	834,129 &	3.64\% \\
         20 to 29 &	6,361,697 &	27.75\% \\
         30 to 39 &	7,357,788 &	32.09\% \\
         40 to 49 &	4,479,057 &	19.54\% \\
         50 to 59 &	2,383,591 &	10.40\% \\
         60 or above	 & 1,509,501 &	6.58\% \\
         Total &	22,925,763 &	100\% \\
        \hline
\end{tabular}}
\label{table6}
\end{table}

When looking at Table \ref{table7}, 73.9\& of participants were full-time employees and 21\% were retired individuals. This is typical since full-time employees do not have the time to visit government agencies to get services physically. It also coincides with responses to the age category question.

\begin{table} [h] 
\caption{Participants' employment status}
\centering
{\begin{tabular}{lcc}
        \hline
        Employment & Count &	Percentage \\
        \hline
        Student &	9 &	3.3\% \\
        Part-time Employee &	1 &	0.4\% \\
        Full-time Employee &	204 &	73.9\% \\
        Looking for a job &	4 &	1.5\% \\
        Retired &	58 &	21\% \\
        Total &	276 &	100\% \\
        \hline
\end{tabular}}
\label{table7}
\end{table}

As for the level of education of participants, Table \ref{table8} shows that 35.9\% hold a bachelor's degree, 33\% hold a master’s degree, and 14.5\% hold a Ph.D. This indicates that the majority of the questionnaire participants were highly educated and that higher levels of education promote e-government usage.

\begin{table} [h] 
\caption{Participants' level of education}
\centering
{\begin{tabular}{lcc}
        \hline
        Education & Count &	Percentage \\
        \hline
        Below high school &	5 &	1.8\% \\
        High school &	18 &	6.5\% \\
        Diploma &	16 &	5.8\% \\
        Some college, but no degree &	7 &	2.5\% \\
        Bachelor &	99 &	35.9\% \\
        Masters &	91 &	33\% \\
        PhD &	40 &	14.5\% \\
        Total &	276 &	100\% \\
        \hline
\end{tabular}}
\label{table8}
\end{table}

Two questions in the questionnaire shed extra light on the importance of the e-government program in Saudi Arabia: “Do you use e-government” and “To what extent are you satisfied with e-government services?” the former shows the popularity of e-government, where the latter indicates the level of satisfaction of e-government. Therefore, additional analysis was done on the relationship between these two questions on one hand, and the participants' basic information on the other, Tables \ref{table11}-\ref{table16} show the results of such relationships.

Analyzing participants' age in Table \ref{table11} and Table \ref{table12} shows that the highest participating age group was 30-39 years old: 63.2\& of this age group were satisfied with e-government and 32.6\% were satisfied to some extent. Also, this age group was ranked as the highest in the usage of e-government (98.9\%).

\begin{table} [h] 
\caption{Participants' age vs. satisfaction with e-government}
\centering
{\begin{tabular}{lcccc}
        \hline
        Age	& Satisfied	& To some extent & Not satisfied & Total \\
        \hline
        Below 20 	     &	2	(100\%)	&	0 (0\%)		& 0 (0\%)		 & 2 \\
        From 20 to 29	 &	8 (44.4\%)	&	9 (50\%)	& 1 (5.6\%)	     & 18 \\
        From 30 to 39	 &	60 (63.2\%)	&	31 (32.6\%)	& 4 (4.2\%)	     & 95 \\
        From 40 to 49	 &	57 (72.2\%) &	20 (25.3\%)	& 2 (2.5\%)	     & 79 \\
        From 50 to 59	 &	29 (63.0\%)	&	17 (37.0\%)	& 0 (0\%)	     & 46 \\
        60 or more	     &	22 (61.1\%) &	14 (38.9\%)	& 0 (0\%)	     & 36 \\
        Total	         &	178	        &	 91	        &	 7 	          & 276 \\
        \hline
\end{tabular}}
\label{table11}
\end{table}

\begin{table} [h] 
\caption{Participants' age vs. e-government usage}
\centering
{\begin{tabular}{lccc}
        \hline
        Age	& Yes	&	No	& Total \\
        \hline
        Below 20	 &	2 (100\%)	&	0 (0\%)	 &	2 \\
        From 20 to 29	 &	16 (88.9\%)	&	2 (11.1\%)	&	18 \\
        From 30 to 39	 &	94 (98.9\%)	&	1 (1.1\%)	&	95 \\
        From 40 to 49	 &	78 (98.7\%)	&	1 (1.3\%)	&	79 \\
        From 50 to 59	 &	46 (100\%)	&	0 (0\%)	     & 	46 \\
        60 or more	     &	36	(100\%)	&	0 (0\%)	 &	36 \\
        Total	 &	272	&	4	&	276 \\
        \hline
\end{tabular}}
\label{table12}
\end{table}

As mentioned above, most participants were full-time employees. Table \ref{table13} and Table \ref{table14} show employment status in regard to both questions: 64.2\& of full-time employees were satisfied, and 33.3\% were satisfied to some extent. Interestingly, 99\& of the same group use e-government.

\begin{table} [h] 
\caption{Participants' employment status vs. satisfaction with e-government}
\centering
{\begin{tabular}{p{.25\linewidth}cccc}
        \hline
        Employment	& Satisfied	& To some extent & Not satisfied &Total \\
        \hline
        Student 	           &	7 (77.8\%)	&	2 (22.2\%)	&	0 (0\%)	 &	9 \\
        Part-time Employee	   &	0 (0\%)	     & 	1 (100\%)	&	0 (0\%)	 &	1 \\
        Full-time Employee	   &	131 (64.2\%)	&	68 (33.3\%)	&	5 (2.5\%)	&	204 \\
        Looking for a job	   &	4 (100\%)	&	0 (0\%)	   &	0 (0\%)	 &	4 \\
        Retired 	           &	36 (62.1\%)	&	20 (34.5\%)	&	2 (3.4\%)	&	58 \\
        Total 	             &	178			 & 91	&	  7	&		  276 \\
        \hline
\end{tabular}}
\label{table13}
\end{table}

\begin{table} [h] 
\caption{Participants' employment status vs. e-government usage}
\centering
{\begin{tabular}{lccc}
        \hline
        Employment	& Yes	&	No & Total \\
        \hline
        Student	 &	8 (88.9\%)	&	1 (11.1\%)	&	9 \\
        Part-time Employee	&	1 (100\%)	&	0 (0\%)		& 1 \\
        Full-time Employee	&	202 (99\%)	&	2 (1\%)	&	204 \\
        Looking for a job	&	4 (100\%)	&	0 (0\%)		& 4 \\
        Retired	 &	57 (98.3\%)	&	1 (1.7\%)	&	58 \\
        Total 	&	272	&		 4	&	 276 \\
        \hline
\end{tabular}}
\label{table14}
\end{table}

The education level was examined in regard to both questions. Bachelor's degree holders were the most satisfied with e-government (67.7\% satisfied and 31.3\% were satisfied to some extent). Additionally, 97\& of bachelor's degree holders acknowledged using e-government, as depicted in Table \ref{table15} and Table \ref{table16}.

\begin{table} [h] 
\caption{Participants' education level vs. satisfaction with e-government}
\centering
{\begin{tabular}{p{.3\linewidth}cccc}
        \hline
        Education	& Satisfied	& To some extent & Not satisfied &Total \\
        \hline
        Below high school 	&	5 (100\%)	&	0 (0\%) 	&	0 (0\%)	&	5 \\
        High school 	&	14 (77.8\%)	&	4 (22.2\%)	&	0 (0\%)	&	18 \\
        Diploma	 &	10 (62.5\%)	&	5 (31.3\%)	&	1 (6.3\%)	&	16 \\
        Some college, but no degree	&	2 (28.67\%)	&	5 (71.4\%)	&	0 (0\%)	&	7 \\
        Bachelor 	&	67 (67.7\%)	&	31 (31.3\%)	&	1 (1.01\%)	&	99 \\
        Masters	 &	56 (61.5\%)	&	32 (35.2\%)	&	3 (3.3\%)	&	91 \\
        PhD 	& 24 (60\%)	 &	14 (35\%)		& 2 (5\%)	&	40 \\
        Total 	&	178	&	  91	&	  7	&	  276 \\
        \hline
\end{tabular}}
\label{table15}
\end{table}

\begin{table} [h] 
\caption{Participants' education level vs. e-government usage}
\centering
{\begin{tabular}{lccccc}
        \hline
        Education	& Yes	&	No & Total \\
        \hline
        Below high school 	&	5 (100\%)	&	0 (0\%) 	&	5 \\
        High school 	&	17 (94.4\%)	&	1 (5.6\%)	&	18 \\
        Diploma 	&	16 (100\%)	&	0 (0\%) 	&	16 \\
        Some college, but no degree	&	7 (100\%)	&	0 (0\%)	&	7 \\
        Bachelor 	&	96 (97\%)	&	3 (3\%)	&	99 \\
        Masters	 &	91 (100\%)	&	0 (0\%) 	&	91 \\
        PhD	 &	40 (100\%)	&	0 (0\%)	 & 40 \\
        Total	 &	272	&		 4	&	  276 \\
        \hline
\end{tabular}}
\label{table16}
\end{table}

In addition, the use of e-government was ranked above 90\%, this indicates that e-government in Saudi Arabia is very popular; it also indicates that the technological infrastructure is adequate, and beneficiaries have the capabilities and knowledge to use e-government.

One of the main questions asked was related to the usage of e-government, 98.55\& of participants agreed that they do, and when they were asked about the frequency of using e-government services 27.2\% said they used it weekly, 26.5\% used it once a month, 20.2\% used it every 2 to 3 months, 19.9\% used it daily, and 6.3\% used it 2 to 3 times a month, as shown in Table \ref{table21} and Table \ref{table22}. This indicates that not all participants need to use e-government services frequently. However, when the e-government service is required, it should be available and within reach. It also encourages e-government service providers to adopt technological solutions that will eliminate any drop or interruption in the provided services.

\begin{table} [h] 
\caption{Participants' use of e-government}
\centering
{\begin{tabular}{lcc}
        \hline
        Usage	& Count	&	Percentage \\
        \hline
        Yes 	&	272	&	98.55\%	\\
        No 	&	4	&	1.45\%	\\
        Total	 &	276	&	100\%	\\
        \hline
\end{tabular}}
\label{table21}
\end{table}

\begin{table} [h] 
\caption{Participants' e-government frequency of use}
\centering
{\begin{tabular}{lcc}
        \hline
        Frequency	 & Count	&	Percentage \\
        \hline
        Daily	 & 54 &	19.9\% \\
        Weekly	 & 74	& 27.2\% \\
        Once a Month	 & 72	& 26.5\% \\
        Every 2-3 Months & 55	& 20.2\% \\
        2-3 Times a Month	& 17	& 6.3\% \\
        Total	 & 272	& 100\% \\
        \hline
\end{tabular}}
\label{table22}
\end{table}

Since there are more than 2300 services provided through e-government in Saudi Arabia \cite{government_of_saudi_arabia_digital_2023}, it is important to know which is the most used service. Therefore, participants were asked about the e-government service they used most frequently: 83.33\& of participants answered “Absher” \cite{saudi_arabian_government_absher_nodate}. Absher is the e-government portal for the Ministry of Interior \cite{saudi_arabian_government_ministry_nodate}. This is anticipated since several citizen-related e-government services are provided through Absher, such as driver's license and passport renewals.

Participants were also asked if they complete their e-government services on their own, and 90.79\% indicated that they did. However, 9.21\% acquire help from either a specialized office, a family member, or others. This may be due to their technological insufficiency, or it could also be the result of not being aware of the resulting risks, which can make them susceptible to illegal attacks or actions. Therefore, e-government service providers need to simplify the user interfaces of their solutions and services, provide training to mitigate technological inadequacy, and intensify awareness campaigns of security threats and attacks.

\subsection{General Satisfaction}

The first aspect is represented by the question “When utilized in the designated context, do e-government services typically meet the needs of beneficiaries?” To answer this question, participants were given five questions that were related to the overall satisfaction of e-government services.

It is noticeable that Saudi Arabian e-government services are very popular, and participants are generally satisfied with them: the majority agreed that e-government services fulfill their requests, their desired service is completed easily, and that online services are convenient. As for satisfaction with e-government services and satisfaction with ease of use, the results are similar, slightly above 60\% are satisfied, and 32\% are satisfied to some extent. These results indicate that e-government services do provide the required services to citizens. In addition, citizens do not need to visit the government organizations' physical office, and they can also complete their requests even after hours. On the other hand, automating services allow government employees to work on other tasks, processes, and functions that require human intervention.

In the context of providing services, request fulfillment is a key element, providing a service to beneficiaries without it is meaningless, thus participants were asked if e-government services fulfill their requests; 86.2\% agreed that they do. In addition, if completing a service is troublesome or not easy, users will resort to finding other means. In the e-government case, users may want to personally visit physical locations of government service providers. Classifying a service as difficult allows for improvement of it for better customer/user satisfaction.

Questionnaire responses show that 90.2\& of participants agreed that e-government services were completed with ease, which seems to look like an excellent achievement. However, 9.8\& of participants do not agree, propagating this to the study's population indicates that approximately two million beneficiaries encounter difficulties when completing e-government services. Therefore, further investigation is required by providers of e-government services. Moreover, participants who answered “no” to this question were asked about the cause of the difficulty, and eight choices were given to select from while giving them the option to select more than one cause: 16.5\% encountered difficulties while using the service, 15.29\% agreed that the service information was not clear, and 14.12\% believed that information provided was not sufficient to meet their needs.

The main objective of providing government services online was to make these services more convenient for beneficiaries, thus participants were asked specifically about the convenience of e-government services in regards to the services provided in person, and 86.2\% agreed that e-government services were more convenient. However, slightly above 11.3\% believed that e-government services were sometimes convenient, and 2.5\% disagreed. This indicates that one-sixth of beneficiaries have doubts about the convenience of e-government services, which require service providers to further investigate the cause individually.

Understanding the extent of both satisfaction and ease of use is important since both play a role in user interface design. Responses showed that 64.5\% and 62.7\& of participants (respectively) were satisfied, while 33\% and 32.6\& of participants, respectively, were satisfied to some extent.

According to the results shown in Table \ref{table17}, it is apparent that most requests were fulfilled since the mean did not exceed 1.14. Most services were also completed with ease, since the mean was 1.1. In addition, most participants believed that online services were convenient, and they were satisfied with e-government services. They were also satisfied with the ease of use of e-government services since the Mean for each question was 1.16, 1.38, and 1.42, respectively.

\begin{table} [h] 
\caption{Participants' response to general satisfaction questions}
\centering
{\begin{tabular}{p{.70\linewidth}ccc}
        \hline
        Question &	Min. &	Max. &	Mean \\
        \hline
        Do e-government services fulfill your requests? &	1.00 &	3.00 &	1.14 \\
        Is the service completed with ease?  &	1.00 &	2.00 &	1.1\\
        The online service was more convenient than the one in person. &	1.00 &	3.00 &	1.16\\
        To what extent are you satisfied with e-government services? &	1.00 &	3.00 & 1.38\\
        To what extent are you satisfied with the e-government service's ease of use? &	1.00 &	3.00 &	1.42 \\
        \hline
\end{tabular}}
\label{table17}
\end{table}

\subsection{Satisfaction with Features}

The second aspect is represented by the question: “Are users satisfied with the features of e-government services?” To study the usefulness of e-government services, participants were given eight questions, as depicted in Table \ref{table18}.

\begin{table} [h] 
\caption{Participants' response to satisfaction with features questions}
\centering
{\begin{tabular}{p {.7\linewidth}ccc}
        \hline
        Question & 	Min. &	Max. &	Mean \\
        \hline
        Was the service you requested fulfilled? &	1.00 &	2.00 &	1.04 \\
        Does service execution consume more time than expected? &	1.00 &	2.00 &	1.86  \\
        Does the service require unnecessary information or steps? &	1.00 &	2.00 &	1.79  \\
        Does the requested service contain an explanation about how it works? &	1.00 &	2.00 &	1.27 \\
        Is the service provided across multiple (different) devices? &	1.00 &	2.00 &	1.16 \\
        To what extent are you satisfied with the look and feel of e-government services? &	1.00 &	3.00 &	1.37 \\
        Is the registration process of e-government services easy? &	1.00 &	3.00 &	1.39  \\
        Is signing into an e-government service is seamless? &	1.00 &	3.00 &	1.26  \\
        \hline
\end{tabular}}
\label{table18}
\end{table}

Feature satisfaction is an essential component when studying user contentment of e-government services. Therefore, participants were asked about eight related questions. These questions allow for the understanding of different aspects of the e-government service.

Request fulfillment indicates that the design and implementation of the service were adequate, the objective of using the service was achieved, and the outcome of the service is what was expected. Of the participants, 96\% agreed that the service they requested was fulfilled.

To standardize and better improve e-government services, DGA requires that each government agency provides a detailed description, an estimated time to complete, and the expected outcomes of each service it provides. In their response, 86.2\% of participants believe that the service did not consume more time than expected, which left 13.8\% to believe that it did. This requires service providers to study the causes of delay (hardware, software, process design, and/or other related causes) and suggest resolutions.

Similarly, 21\% of participants agree that services require unnecessary information or steps. Such an agreement, to a high percentage, indicates that there exists a problem with service design that needs to be resolved. In addition, 27.2\% of participants did not agree that there is an explanation about the service's nature of work. This indicates that the learning curve to fully understand how a service works may be extendable, and the promising benefit of using e-government may not be fulfilled. Furthermore, this also indicates that some government agencies exist that do not fully comply with DGA's requirements.

The availability of e-government services allows for completion of its goals. Allowing access to the service from several devices enhances its availability. Among the participants, 83.7\% agreed that e-government services are available across multiple devices. However, 16.3\% do not, which raises the issue of lack of awareness. Thus, e-government service providers need to properly promote and advertise methods and devices used to access their services.

Designing the user interface of automated services is essential when it comes to its acceptance. The interaction between the user and the service also needs to be designed using state-of-the-art techniques. Therefore, participants were asked three questions related to the user interface and the interaction with the service, “To what extent are you satisfied with the look and feel of e-government services?” for the former, and “Is registration for e-government services easy” and “Is signing into an e-government service seamless” for the latter. The registration and the login processes were chosen since all e-government services require these two processes. In response to satisfaction with the look and feel of the service, 65.6\% were satisfied, and 31.5\% were satisfied to some extent. This constitutes revisiting the user interface design by the service providers. As for participants' agreement on the ease of registration and logging in, results show that agreement reached 67\% and 77.5\%, respectively. On the other hand, 26.8\% and 18.8\%, respectively, agreed to some extent, and 6.2\% and 3.6\%, respectively, disagreed, which points to issues with the interaction between users and these services. This necessitates that service providers investigate causes and provide solutions for such issues.

In this aspect, participants answered the first five questions with yes or no. In the sixth question, they answered with satisfied, satisfied to some extent, or not satisfied. As for the last two questions, answers were either yes, to some extent, or no.

The majority of participants agreed that the requested service was fulfilled, it contained an explanation about its features, and it was provided across multiple devices. These conclusions could be drawn based on the mean value of each corresponding question (1.04, 1.27, and 1.16, respectively), which is close to 1 (response option: yes). In addition, participants also agreed that the service execution does not consume more time, nor does it require unnecessary information or steps, this is apparent from the mean value for both questions, 1.86 and 1.79, respectively, which is closer to 2 (response option: no).

In addition, participants were generally satisfied with the look and feel of the e-government services. This was clarified by examining the value of the Mean (1.37), which was close to 1 (response option: satisfied). Furthermore, the majority of participants agreed that the registration was easy, and signing into the e-government service was seamless. This was apparent from the value of the mean, which was 1.39 and 1.26, respectively, close to 1 (response option: yes).

Questions in this part were closely related to the user interface and/or the processes and procedures to complete any given task. It is advised that e-government service providers adopt state-of-the-art techniques to improve the user interface. In addition, it is advised that processes and procedures are reviewed for better automation.

\subsection{User Trust}

The third aspect was represented by the question: “Are e-government services trusted by users?” To study the user's trust in e-government services, question 21 was included in the questionnaire. Participants were asked if they agreed or disagreed with six specific statements related to trust by answering yes or no to each statement.

Trustworthiness is an important aspect of any product. When it comes to software, no user is interested in using the software unless they are assured it is worth being trusted, especially when personal and/or financial data and information are involved. Therefore, one of the questions was specifically linked to trustworthiness. In addition, having any suspicions in the outcome of any service is troublesome, and may lead to not using the service at all. Hence, participants were asked if they have any suspicion of the e-government service results. Moreover, having any doubts that a service could cause its users harm is problematic: users must be assured that e-government services will not cause them harm, especially if the safety of their personal and/or financial information is uncertain. Thus, a question relating to the participants' feeling of security was included in the questionnaire. Finally, participants were asked about their confidence in e-government services, their familiarity with e-government services, and if e-government services were completed successfully. All three questions enabled judging whether using e-government services causes any serious trust issues or problems.

Most participants believed that the e-government services were trustworthy, they had confidence in them, they were familiar with them, the e-government services were completed successfully, they were not suspicious of the e-government services, nor did they believe they caused harm. This is shown in Table \ref{table19}, where the Mean of answers to these questions is close to 1 (response option: yes). 

\begin{table} [h] 
\caption{Participants' response to trust questions}
\centering
{\begin{tabular}{p{.60\linewidth}ccc}
        \hline
        Question & 	Min. &	Max. &	Mean \\
        \hline
        The e-government service is trustworthy &	1.00 &	2.00 &	1.03 \\
        I am not suspicious of the e-government service's results &	1.00 &	2.00 &	1.11 \\
        The e-government service's actions do not cause harm &	1.00 &	2.00 &	1.05 \\
        I am confident in the e-government service &	1.00 &	2.00 &	1.03 \\
        I am familiar with the e-government service &	1.00 &	2.00 &	1.06 \\
        The e-government service is completed successfully &	1.00 &	2.00 &	1.04 \\
        \hline
\end{tabular}}
\label{table19}
\end{table}

Furthermore, responses to these questions point to a high level of trust in e-government services (89\% or more). This indicates that users were well aware of the security status of e-government services and were constantly informed of security threats. Participants acknowledge that their confidence in e-government services is very high (96.7\%). Responses to questions in this part signified that a high level of user trust existed and there was confidence in the e-government services provided in Saudi Arabia.

\subsection{Enjoyment (Pleasure)}

The fourth and final aspect was represented by the question: “Do consumers find e-government services enjoyable to use?” Therefore, question 22 was added to allow study of the level of pleasure users obtain when using e-government services. Participants were asked about eight specific feelings while allowing them one of three responses for an answer: not at all, moderately, or completely.

Quantifying feelings overall is known to be difficult. Therefore, participants were explicitly asked about their individual feelings, while describing the specific circumstances that surround them as they assessed their feelings when answering the question. Participants were asked to assess their excitement, enthusiasm, pride, distress, carelessness, shame, jitteriness, and fear. Such assessment allowed for understanding of the amount of pleasure gained when participants use e-government services. It is noticeable that the list of feelings were both positive and negative. The first three questions were related to feelings that were considered positive, while the remaining questions were related to feelings that are known to be negative.

To fully obtain a full pleasure score for using e-government services, participants needed to answer the first three questions with “completely,” and the last five with “not at all.” The Mean of the first three was close to 3 (response option: completely); this implied that the majority of participants were excited, enthusiastic, and proud when they used the e-government service. On the other hand, the Mean of the last five feelings was closer to 1 (response option: not at all), which indicates that the majority of participants did not feel distressed, careless, ashamed, jittery, or afraid when using the e-government service. Table \ref{table20} presents these results.

\begin{table} [h] 
\caption{Participants' response to pleasure questions}
\centering
{\begin{tabular}{lccc}
        \hline
        Question & 	Min. &	Max. &	Mean \\
        \hline
        Excited &	1.00 &	3.00 &	2.28 \\
        Enthusiastic &	1.00 &	3.00 &	2.3 \\
        Proud &	1.00 &	3.00 &	2.55 \\
        Distressed &	1.00 &	3.00 &	1.26 \\
        Careless &	1.00 &	3.00 &	1.31 \\
        Ashamed &	1.00 &	2.00 &	1.18 \\
        Jittery &	1.00 &	3.00 &	1.25 \\
        Afraid &	1.00 &	3.00 &	1.25 \\
        \hline
\end{tabular}}
\label{table20}
\end{table}

It is worth noting that the mean for the “proud” feeling was 2.55; this is a positive indicator of the high sense of pride from participants in the work done by the Saudi Arabian government in its e-government program.

\subsection{Discussion: Satisfaction, Trust, and the Nature of Digital Citizenship in Saudi Arabia}

The descriptive analysis indicates a population that is exceptionally skilled in utilizing and generally content with the functional results of Saudi e-government services. This high level of transactional satisfaction (as shown in Table \ref{table17}, where the mean scores for fulfillment and ease were close to 1.0) is strongly correlated with the kingdom's elite standing in global e-government rankings and reflects the significant investment made through its Vision 2030 digital transformation initiative. Nevertheless, a more profound and critical examination of the data using this works integrated ISO/UTAUT framework uncovers subtle tensions and suggests a more intricate relationship between digital citizens and the state.

The elevated ratings for usability and functional transparency (for instance, 90.2\% of users found the services easy to complete, Table \ref{table17}) are in strong accordance with the UTAUT construct of Effort Expectancy. Users perceive the systems clear and easy to use, which diminishes the cognitive obstacles to adoption. Similarly, the perceived reliability and successful task completion (96\% reporting service fulfillment, Table \ref{table18}) bolster Performance Expectancy; the systems are regarded as effective tools for accomplishing specific objectives. This phenomenon elucidates the high adoption rates and overall satisfaction, especially among employed individuals who are digitally literate. However, this functional success sharply contrasts with the significantly lower levels of emotional engagement (Pleasure Means ranging from 1.18 to 2.55, Table \ref{table20}). This suggests a digital government model that excels at governing but less effective at engaging; the government entities are efficient service providers but not yet an engaging digital space.

The data highlights a pivotal aspect of Saudi Arabia's digital transformation: the integration of the digital citizen experience. This is exemplified by the significant prevalence of the "Absher" platform, 230 of 276 participants (83.33\%) named “Absher” as their most frequently used e-government service. Its widespread use is not simply a matter of user choice but stems from a well-planned, cohesive governance framework. By functioning under a central organization that oversees a comprehensive network of over 20 essential government entities (civil affairs, traffic, passports, the national single sign-on service (Nafath), etc.) Absher serves as a consolidated entry point to the government services. This architecture promotes digital citizenship by providing unmatched convenience and coherence, effectively integrating disparate services into a singular, recognizable user journey. It fortifies digital citizenship by establishing a foundation of trust through consistent performance, proven reliability, and the robust security protocols embedded in its operational DNA. As a result, Absher evolves beyond being a simple portal; it has emerged as the primary, reliable environment where digital citizenship is not only facilitated but actively practiced and experienced. The exceptionally high user confidence ratings (e.g., 96.7\%) are a direct validation of this successful, centralized platform-of-platforms approach.

Moreover, the challenges reported (21\% identified unnecessary steps, 27.2\% perceived lack of explanations, Table \ref{table18}) indicate a disconnect between technical functionalities and user-centered design. This issue transcends mere usability; it pertains to digital equity. The findings imply that individuals who need assistance in accessing services (9.21\% of respondents) or who find these services inconvenient may experience these obstacles as particularly exclusionary. Consequently, this may create a rift between those who can easily navigate the digital landscape and those who struggle to do so (as suggested by the demographic differences in Tables \ref{table11}, \ref{table13}, and \ref{table15}), thereby compromising the inclusive objectives of e-government.

For Saudi Arabia's digital transformation to cultivate a more dynamic digital public sphere, future designs should integrate emotional design principles that transcend simple transactions. The results indicate that enhancing functional clarity and inclusive design is equally essential as ensuring robust security and uptime to establish a genuinely comprehensive and engaging digital citizenship.


\section{Conclusion and Prospective}
\label{Conclusion}

This research evaluated the satisfaction levels of Saudi citizens regarding e-government services through a detailed questionnaire based on the ISO/IEC 25010 QinU framework. By concentrating on four primary dimensions - overall satisfaction, satisfaction with features, trust, and enjoyment - the study provides a systematic insight into user experiences with digital government platforms.

The overall findings suggest that e-government services in Saudi Arabia are extensively utilized and generally well-received. Participants consistently reported high levels of trust, ease of use, and successful service completion. These outcomes correspond with Saudi Arabia’s prominent position in the 2024 UN E-Government Development Index (EGDI), further validating the effectiveness of its digital transformation efforts.

Saudi Arabia's impressive satisfaction ratings (for instance, 86.2\% in service fulfillment) reflect its elite position in the EGDI rankings (6th worldwide in 2024 \cite{united_nations_department_of_economic_and_social_affairs_united_2024}) and correspond with trends seen in comparable 'Very High EGDI' countries such as the UAE (0.9568 EGDI) and South Korea (0.9630 EGDI) \cite{united_nations_department_of_economic_and_social_affairs_united_2024}. These nations also emphasize usability and trust in their e-government services, indicating that Saudi Arabia's digital transformation approach is in harmony with international best practices for citizen-focused design.

Nevertheless, the analysis also revealed areas needing enhancement. Some users indicated challenges with service clarity, excessive information requests, and limited compatibility across devices. Additionally, a significant number of users expressed only moderate satisfaction with the design of user interfaces and features.

\subsection{Implications for Practice}

The findings present several practical implications for government agencies, policymakers, and service designers:
\begin{enumerate}
    \item Trust and Transparency: It is essential to enhance user trust. Clear communication, visible privacy guarantees, and transparent service processes can further bolster confidence in digital government.
    \item User-Centered Design: Enhancing clarity, consistency, and visual appeal of service interfaces may contribute to higher satisfaction levels. Agencies should prioritize user experience (UX) research and ongoing usability testing.
    \item Accessibility and Awareness: Although overall usage is high, awareness initiatives should focus on underrepresented or hesitant user groups, particularly older citizens, to foster inclusivity.
    \item Data-Driven Monitoring: Merging system usage analytics with satisfaction surveys can offer a more comprehensive perspective on service performance and user behavior.
\end{enumerate}

\subsection{Limitations}

While this research offers significant insights into citizen satisfaction with e-government services in Saudi Arabia, it is essential to recognize several limitations.

First, all data were self-reported, which may introduce biases such as social desirability or inaccuracies in recall. Respondents might have provided favorable responses due to perceived connections with national identity or apprehension about potential consequences. Triangulating self-reported data with objective metrics (e.g., system logs, usability assessments) and qualitative interviews could enhance the reliability of the findings.

Second, although the questionnaire was constructed using the ISO/IEC 25010 and 25022 QinU models, it was tailored specifically for the Saudi e-government context. Thus, caution is warranted when applying it to different countries or cultures without appropriate localization, translation, and revalidation.

Third, while the “pleasure” construct introduces a human-centered aspect to the evaluation, the emotional metrics employed in this study were not grounded in established affective or user engagement frameworks. Future investigations should utilize validated tools such as PANAS, the User Engagement Scale, or the Self-Assessment Manikin (SAM) to ensure enhanced construct validity.

Finally, the study relied solely on descriptive statistics for result analysis. No inferential or multivariate analyses were performed to identify significant predictors of satisfaction. Future studies should integrate statistical methodologies such as regression analysis or structural equation modeling to explore causal or correlational relationships among variables.

\subsection{Future Work}

Future investigations should expand upon this research in multiple avenues:
\begin{enumerate}
    \item Integrate inferential statistical techniques to reveal connections between demographic factors and satisfaction results.
    \item Widen the range of the survey to encompass additional ISO/IEC 25010 QinU characteristics such as effectiveness, efficiency, and context coverage.
    \item Modify and validate the survey for application in various cultural or national settings.
    \item Perform targeted assessments of significant platforms like "Absher" utilizing task-oriented evaluations and user journey analysis.
    \item Investigate service-specific categories (e.g., G2C, G2B, G2G) to pinpoint domain-specific quality requirements and satisfaction influences.
    \item Investigate regional inequalities, given that users in rural areas may encounter difficulties with connectivity.
    \item Although this research is consistent with the constructs of UTAUT, subsequent studies might explicitly evaluate the complete model using structural equation modeling.
\end{enumerate}

\subsection{Practical Recommendations for E-Government Designers}

Based on the research findings, the following design-focused recommendations are proposed to enhance user satisfaction:
\begin{enumerate}
    \item Streamline Form Fields: Reduce unnecessary or overly complicated input requirements to alleviate user frustration.
    \item Clarify Service Communication: Employ straightforward language and provide guided instructions to clarify the purpose, steps, and anticipated outcomes of each service.
    \item Improve Mobile Usability: Enhance responsiveness and interactivity for mobile users, who have reported a higher incidence of usability challenges.
    \item Establish Feedback Mechanisms: Offer users real-time status updates or confirmation messages to bolster their confidence in the success of the service.
    \item Assist First-Time Users: Provide onboarding tutorials or detailed walkthroughs to mitigate anxiety and enhance initial user experiences.
\end{enumerate}

\section*{Statements and Declarations}
\label{statements}

\begin{description}
    \item[Author contribution] The Author is responsible for the conceptualization of this research work. He developed the questionnaire, prepared the figures of the manuscript, wrote the main manuscript, and reviewed the manuscript draft.
    \item [Data Availability] The author declares that the data used in this study are available in the article. In addition, detailed questionnaire responses are available from the corresponding author on reasonable request after removing any sensitive information based on data privacy laws in the Kingdom of Saudi Arabia.
    \item [Competing interests] The author declares that there are no competing interests.
    \item [Funding] No funds, grants, or other support was received for this work.
    \item [Conflict of interest] The author certifies that he has no affiliations or involvement in any organization or entity with any financial or non-financial interest in the subject matter or materials discussed in this manuscript.
    \item [Ethical Considerations] Participation in this research was both voluntary and anonymous. Participants were made aware of the study's objectives and were guaranteed that no personally identifiable information would be gathered. The research did not include any vulnerable groups. Informed consent was secured from all participants prior to the administration of the survey, and all responses were stored and analyzed in compliance with ethical standards for research involving human subjects.
\end{description}


\appendix

\section{Questionnaire}
\label{Questionnair}
Below is a copy of the questionnaire, complete with all potential answers, that was given to participants surveyed in this study.

\begin{center}
    \textbf{Basic information}
\end{center} 

\begin{enumerate}
    \item In which age category do you belong?
        \begin{itemize}
            \item Below 20
            \item From 20 to 29
            \item From 30 to 39
            \item From 40 to 49
            \item From 50 to 59
            \item 60 years or older
        \end{itemize}
    
    \item Which of the following categories best describes your employment status?
        \begin{itemize}
            \item Student
            \item Part-time Employee
            \item Full-time Employee
            \item Looking for a job
            \item Retired
        \end{itemize}
    
    \item What is your level of education?
        \begin{itemize}
            \item Below high school
            \item High school
            \item Diploma
            \item Some college, but no degree
            \item Bachelor
            \item Masters
            \item PhD
        \end{itemize}
    
    \begin{center}
        \textbf{E-government services and beneficiary needs}
    \end{center} 
    
    \item Do you use e-government?
        \begin{itemize}
            \item Yes (an extension to the question will be displayed)
            \item No
        \end{itemize}
    How often do you use e-government services:
        \begin{itemize}
            \item Daily
            \item Weekly
            \item Once a Month
            \item Every 2-3 Months
            \item 2-3 Times a Month
        \end{itemize}
    
    \item Do e-government services fulfill your requests?
        \begin{itemize}
            \item Yes
            \item To some extent
            \item No
        \end{itemize}
    
    \item Which e-government service do you use the most?
        \begin{itemize}
            \item Textual feedback
        \end{itemize}
    
    \item Was the service you requested fulfilled?
        \begin{itemize}
            \item Yes
            \item No
        \end{itemize}
    
    \item Does service execution consume more time than expected?
        \begin{itemize}
            \item Yes
            \item No
        \end{itemize}
    
    \item Does the service require unnecessary information or steps?
        \begin{itemize}
            \item Yes
            \item No
        \end{itemize}
    
    \item Does the requested service contain an explanation about how it works?
        \begin{itemize}
            \item Yes
            \item No
        \end{itemize}
    
    \item Is the service provided across multiple (different) devices?
        \begin{itemize}
            \item Yes
            \item No
        \end{itemize}
    
    \item Is the service completed with ease?
        \begin{itemize}
            \item Yes
            \item No (an extension to the question will be displayed)
        \end{itemize}
    
    Please select the cause of difficulty (you can select more than one).
        \begin{itemize}
            \item The website was not visually appealing.
            \item I encountered difficulties while using the service.
            \item The services were not completed quickly.
            \item The service cannot be used at any time.
            \item The pages did not load properly.
            \item The information provided is not sufficient to meet my needs.
            \item The information regarding the service was not relevant to me.
            \item Service information is not clear and easy to understand.
        \end{itemize}
    
    \item Do you complete e-government services on your own?
        \begin{itemize}
            \item Yes
            \item No (an extension to the question will be displayed)
        \end{itemize}
    
    To complete a service I get help from:
        \begin{itemize}
            \item A specialized office
            \item Family member
            \item A friend
            \item Other
        \end{itemize}
        
    \begin{center}
        \textbf{Satisfaction with the features of e-government services}
    \end{center}
    
    \item The online service was more convenient than the one in person.
        \begin{itemize}
            \item Yes
            \item Sometimes
            \item No
        \end{itemize}
    
    \item To what extent are you satisfied with e-government services?
        \begin{itemize}
            \item Satisfied
            \item To some extent
            \item Not satisfied
        \end{itemize}
    
    \item To what extent are you satisfied with the e-government service's ease of use?
        \begin{itemize}
            \item Satisfied
            \item To some extent
            \item Not satisfied
        \end{itemize}
    
    \item To what extent are you satisfied with the look and feel of e-government services?
        \begin{itemize}
            \item Satisfied
            \item To some extent
            \item Not satisfied
        \end{itemize}
    
    \item Is the registration process of e-government services easy?
        \begin{itemize}
            \item Yes
            \item To some extent
            \item No
        \end{itemize}
    
    \item Is signing into an e-government service seamless?
        \begin{itemize}
            \item Yes
            \item To some extent
            \item No
        \end{itemize}
    
    \item Do you agree with the following statements:

    \begin{enumerate}
        \item The e-government service is trustworthy.	 
            \begin{itemize}
                \item Yes
                \item No
            \end{itemize}
        \item I am not suspicious of the e-government service's results.
            \begin{itemize}
                \item Yes
                \item No
            \end{itemize}
        \item The e-government service's actions do not cause harm.
            \begin{itemize}
                \item Yes
                \item No
            \end{itemize}
        \item I am confident in the e-government service.
            \begin{itemize}
                \item Yes
                \item No
            \end{itemize}
        \item I am familiar with the e-government service.
            \begin{itemize}
                \item Yes
                \item No
            \end{itemize}
        \item The e-government service is completed successfully.
            \begin{itemize}
                \item Yes
                \item No
            \end{itemize}
    \end{enumerate}

    \item Generally, when signing into an e-government service, indicate your feelings towards each word or term:
\begin{multicols}{2}
    
    \begin{enumerate}
        \item Excited
            \begin{itemize}
                \item Not at all
                \item Moderately
                \item Completely
            \end{itemize}
        \item Enthusiastic
            \begin{itemize}
                \item Not at all
                \item Moderately
                \item Completely
            \end{itemize}
        \item Proud
            \begin{itemize}
                \item Not at all
                \item Moderately
                \item Completely
            \end{itemize}
        \item Distressed
            \begin{itemize}
                \item Not at all
                \item Moderately
                \item Completely
            \end{itemize}
\columnbreak
        \item Careless
            \begin{itemize}
                \item Not at all
                \item Moderately
                \item Completely
            \end{itemize}
        \item Ashamed
            \begin{itemize}
                \item Not at all
                \item Moderately
                \item Completely
            \end{itemize}
        \item Jittery
            \begin{itemize}
                \item Not at all
                \item Moderately
                \item Completely
            \end{itemize}
        \item Afraid
            \begin{itemize}
                \item Not at all
                \item Moderately
                \item Completely
            \end{itemize}
    \end{enumerate}
\end{multicols}{}

\end{enumerate}

  \bibliographystyle{elsarticle-harv} 
  \bibliography{Biblio}

\end{document}